\documentclass[aps,prl,twocolumn,amsmath,amssymb,
superscriptaddress,floatfix,showpacs]{revtex4}

\usepackage{graphicx}
\usepackage{bm}

\begin{document}

\title{Octupolar order in the multiple spin exchange
model on a triangular lattice}

\author{Tsutomu~Momoi}
\affiliation{Condensed Matter Theory Laboratory, RIKEN, Wako,
Saitama 351-0198, Japan}
\author{Philippe~Sindzingre}
\affiliation{
Laboratoire de Physique Th\'eorique de la Mati\`ere Condens\'ee, UMR
7600 of CNRS,  \protect\mbox{Universit\'e P. et M. Curie, case 121,
4 Place Jussieu, 75252 Paris Cedex, France}}
\author{Nic~Shannon}
\affiliation{H. H. Wills Physics Laboratory, University of Bristol,
Tyndall Ave, BS8-1TL, UK}
\date{24 October 2006}

\begin{abstract}
We show how a gapless spin liquid with hidden octupolar order
arises in applied magnetic field, in a model applicable to thin
films of $^3$He with competing ferromagnetic and antiferromagnetic
(cyclic) exchange interactions.  Evidence is also presented for
nematic --- i.e. quadrupolar --- correlations bordering on ferromagnetism
in the absence of magnetic field.
\end{abstract}

\maketitle

The triangular lattice occupies a special place in the history of
quantum magnetism. Although it is the simplest lattice to exhibit
geometrical frustration, this still has highly nontrivial
consequences  --- notably,  that the Ising antiferromagnet (AF) on a
triangular lattice remains disordered even in the limit of zero
temperature. This fact led Anderson to suggest that the ground state
of the spin-1/2 Heisenberg AF on a triangular lattice would be a
gapless spin-liquid state with ``resonating valence bond'' (RVB)
character~\cite{Anderson}. While this conjecture has since been
disproved~\cite{bernu92}, the idea of the RVB state continues to
exert a strong hold on the popular imagination, and has motivated a
search for spin-liquid states in a wide variety of two-dimensional
spin models~\cite{MisguichL}.

Lately, attention has been refocused on the triangular lattice by
the discovery of a number of experimental realizations of a quasi-2D
spin-1/2 magnets with triangular coordination. These include the
organic (BEDTÐTTF)$_2$Cu$_2$(CN)$_3$~\cite{shimizu03}, the
transition metal chloride Cs$_2$CuCl$_4$~\cite{coldea01}, and solid
films of $^3$He absorbed on graphite~\cite{greywall89}.   Of these,
$^3$He films offer the most perfect realization of a spin-1/2
triangular lattice, but with an interesting twist --- the
nearest--neighbor interaction is {\it ferromagnetic} (FM) and
competes with an AF 4-spin cyclic exchange~\cite{roger90}. As such,
$^3$He on graphite is one of a number of new quasi-two dimensional
systems to exhibit ``frustrated
ferromagnetism''~\cite{dresden,kageyama}. However it is unique in
both its purity, and in the possibility of tuning the ratio of the
competing interactions continuously simply by varying the density of
$^3$He atoms. An additional strong motivation for understanding this
system comes from the fact that, for a range of densities bordering
on ferromagnetism, the magnetic ground state of $^3$He films is a
fully gapless spin-liquid~\cite{Fukuyama,Ishimoto}.

Recently, we proposed that the competition between FM and AF
interactions in frustrated spin systems could lead to a new kind of
gapless spin-liquid state with hidden multipolar order bordering on
the FM phase~\cite{shannon04,momoi05,shannon06} (for related,
earlier work see~\cite{andreev84}). In this Letter we extend these
ideas to a triangular lattice spin-1/2 model relevant to $^3$He on
graphite~\cite{roger90}
\begin{equation} {\cal
H}= J \sum_{\langle ij \rangle} P_2 + K \sum_{\langle ijkl \rangle}
(P_4+P_4^{-1}) - h \sum_i S^z_i, \label{Hamiltonian}
\end{equation}
where $P_2$ and $P_4$ (cyclically) permute spins on nearest neighbor
bonds $\langle ij \rangle$ and diamond plaquettes $\langle ijkl
\rangle$, respectively. We include a coupling to magnetic field $h$,
and consider exclusively FM $J <0$ and AF $K>0$.

\begin{figure}[tb]
  \centering
  \includegraphics[width=8.2truecm]{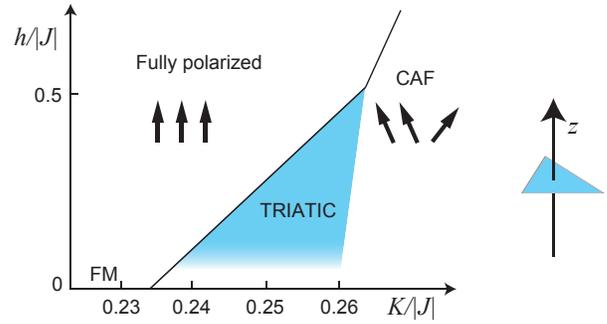}
\caption{(Color online). (Left) Schematic phase diagram for the
spin-1/2 $J$-$K$ multiple spin exchange model on a triangular
lattice in applied magnetic field, in the parameter range where
\protect\mbox{$J<0$} is FM and $K \ge 0$ is AF. A "triatic" phase
with magnetic octupolar order appears in the shaded region between
fully polarized and canted AF (CAF) phases.
 (Right) Schematic spin structure of this triatic order, which is
invariant under spin rotations of $2\pi/3$ about the $z$ axis.
 \label{fig1}}
\end{figure}

Our main finding is that a dynamical process unique to 4-spin
exchange on the triangular lattice leads to the formation of three
magnon bound states which --- under applied magnetic field
--- condense, giving rise to a new, partially-polarized quantum
phase in which rotational symmetry is spontaneously broken, but
without any long-range spin order perpendicular to the field
(Fig.~\ref{fig1}). We identify the order parameter for this novel
phase as a rank three tensor with octupolar character, and confirm
its existence through the exact diagonalization of finite-size
clusters.

To better understand the origins of this new quantum phase, let us
first reconsider the classical $S \to \infty$ limit of
Eq.~(\ref{Hamiltonian}). For $K/|J|<1/4$, the ground state is a
saturated FM. For $1/4<K/|J|<1$ it is highly
degenerate~\cite{momoi97}. Freely propagating domain walls destroy
long range spin order, but nematic (i.e. quadrupolar) order
persists~\cite{momoi05}. This pathology is also evident in the spin
excitations for the quantum spin-1/2 case. Within the FM phase, the
one--magnon dispersion $\omega ({\bf q}) = -(J+4K)
   \left[3 -\cos (q_x) - 2\cos(q_x)
   \cos(\sqrt{3}q_y/ 2)\right]\nonumber$
is well defined. However exactly at the classical critical point $J
= -4K$ it {\it vanishes} identically.

Nondispersing --- i.e. localized --- magnons are also found at the
border of a FM phase in the equivalent multiple spin exchange model
on the square lattice.    In this case, pairs of flipped spins can
propagate coherently under the action of AF spin exchange, even
where a single magnon is localized. Magnons therefore form kinetic
energy driven bound states, which condense and give rise to a
quadrupolar ``bond--nematic'' order~\cite{shannon06}. The energy of
two flipped spins in a FM background can also be calculated {\it
exactly} for the multiple spin exchange model on a triangular
lattice. We find that two-magnon bounds states with total spin $S =
N/2 -2$ and momenta ${\bf q} = \{(0,2\pi/\sqrt{3}),
(\pi,\pi/\sqrt{3}),(-\pi,\pi/\sqrt{3})$\} have a lower energy per
spin than independent magnons for $K/|J|> 0.2349$.
\begin{figure}[tb]
\begin{center}
    \includegraphics[width=58mm]{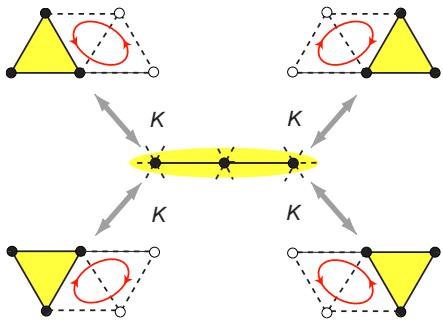}
\caption{(Color online). The four-spin cyclic exchange operator can
be used to propagate three flipped spins ($\bullet$) in a polarized
FM background by tunneling between compact triangular and straight
line configurations.  } \label{fig:triad_hopping}
\end{center}
\end{figure}

However, the frustrated geometry of the triangular lattice introduces a new,
two-step process by which a group of three flipped spins can move together
--- as shown in Fig.~\ref{fig:triad_hopping}. This motivates us to try a simple
trial wave function for a three spin bound state, within the
restricted basis of states where the three spins form a compact
straight line or triangle. Solving the Hamiltonian
Eq.~(\ref{Hamiltonian}) in this restricted Hilbert space, we obtain
five eigenmodes for the three-magnon bound states. The lowest
eigenvalue, $E = -13J - 50 K - \sqrt{J^2 + 4 J K + 28 K^2}$,
corresponds to a state with momentum ${\bf q}=(0,0)$, belonging to
the trivial irrep of the space group, i.e. even under the space
rotations $R_{2\pi/3}$, $R_{\pi}$ and the reflection $\sigma$. This
variational estimate of the three--magnon bound state is lower in
energy than any single magnon or two--magnon state for a range of
parameters $0.2347<K/|J|<0.2652$.

To put these calculations in a more physical context, let us
consider the case in which a magnetic field $h$ is used to fully
polarize the system. This field is then reduced until a critical
value $h_c$ is reached, at which point the magnetization starts to
change. We can calculate critical fields $h_{c1}$ and $h_{c2}$ at
which one- and two-magnons start to Bose-condense. We can also
estimate --- using the trial wave function discussed above --- the
field $h_{c3}$ at which the system becomes unstable against the
condensation of three-magnon bound states. In the absence of any
other phase transition, the greatest of the three fields $h_{c1}$,
$h_{c2}$ and $h_{c3}$ will determine the first instability of the
system, and the nature of magnetic order near to saturation.

\begin{figure}[tb]
\begin{center}
    \includegraphics[width=70mm]{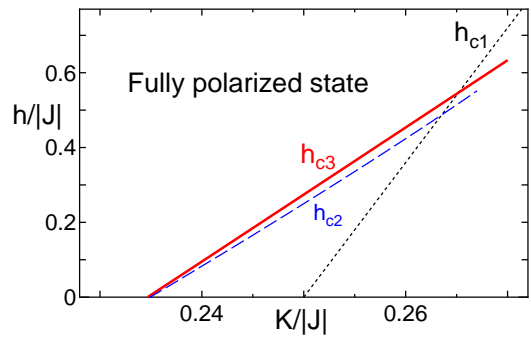}
\caption{(Color online). Critical fields $h_{c1}$ (dotted line),
$h_{c2}$ (dashed line), and $h_{c3}$ (solid line), where one-, two-,
and three-magnon instabilities destabilize the fully polarized state
in the $J$-$K$ model with FM $J<0$. The lines of $h_{c1}$  and
$h_{c2}$ are given from exact solutions. The variationally estimated
line $h_{c3}$ gives an exact lower bound for three-magnon
instability line. } \label{fig:instability}
\end{center}
\end{figure}

Performing this analysis, we find that the first instability for
$K/|J| \gtrsim 0.27$ occurs in the one--magnon channel, and is
against a three-sublattice canted AF, as previously found in mean
field analysis~\cite{momoi97}. However in the parameter range
$0.2347 < K/|J| \lesssim 0.27$, $h_{c3} > h_{c2} \gg h_{c1}$ --- see
Fig.~\ref{fig:instability}. Since the estimate of $h_{c3}$ is
variational (while the values of $h_{c2}$ and $h_{c1}$ are exact),
it provides a strict lower bound on $h_c$.

\begin{figure}[tb]
\begin{center}
    \includegraphics[width=80mm]{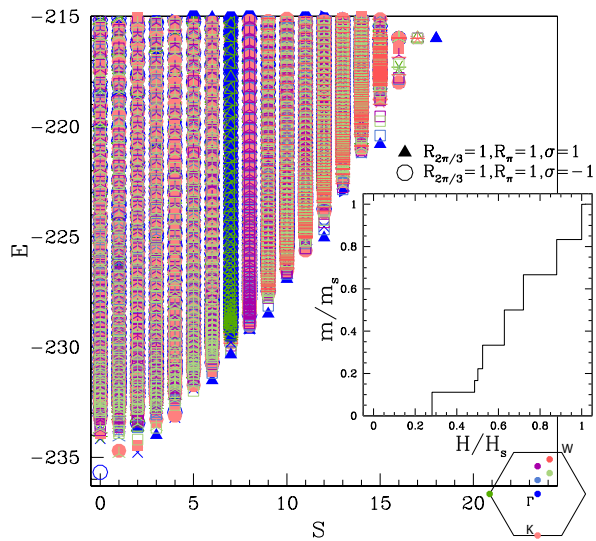}
\caption{Energy spectrum of the $J$-$K$ model for $N=36$ spin
  cluster with $J = -4$ and $K=1$, showing the oscillation in energies of
  the lowest lying states in spin sectors in the period three for large $S$.
  The inset shows the magnetization process $m/m_s$. Jumps of $|\Delta m| = 3$,
  corresponding to octupolar order, are clearly visible at large H. (Color online).}
\label{fig:spectrum}
\end{center}
\end{figure}

Exact diagonalization of finite clusters of up to $N=36$ spins
confirms that the transition out of the saturated state is
controlled by three--magnon bound states for $0.24 \lesssim K/|J|
\lesssim 0.28$.  Results for a 36--spin cluster with $K/|J| = 0.25$
are shown in Fig.~\ref{fig:spectrum}. Jumps of $|\Delta m| = 3$ are
clearly visible in the magnetization process (inset) for fields
close to saturation ($H_s \equiv h_{c3}$).    The strong AF coupling
$K$ ensure that interactions between three--magnon bound states are
repulsive, and excitations containing more than three magnons
therefore play no part in the transition. At finite $H_s$,  the
transition is of second order, and can be understood as a
Bose--Einstein condensation of three--magnon bound states. For $H_s
\to 0$, this gives way to a first order transition.

These jumps can be traced to a set of states which lie at the bottom
of spin sectors $S =3n$, belonging to the trivial irrep of the space
group. Each jump of \mbox{$\Delta m = -3$} corresponds to adding one
more three--magnon bound state to the system.  The convex locus of
these states as a function of $S$ implies that the transition at
$h_{c3}$ is of second order.  Numerical and analytic results close
to saturation are therefore in perfect agreement. The condensation
of three--magnon bound states will lead to a new form of quantum
order. But what kind of order is it ?

We consider first the situation in applied magnetic field, where
full $SU(2)$ spin symmetry is broken down to $U(1)$ rotations about
the z-axis. In this case the order parameter is determined by the
operator $O^\dagger_{ijk}=S_i^- S_j^- S_k^-$, which creates a
three--magnon bound state in a polarized background. Both
$\mathcal{R}e$~$O^\dagger$ and $\mathcal{I}m$~$O^\dagger$ are linear
combinations of fully symmetrized rank-3 tensor operators
\begin{eqnarray}
O^{\alpha\beta\gamma}_{ijk} &=&\sum_{a,b,c \in
P(\alpha,\beta,\gamma)}S_i^a S_j^b S_k^c,
\end{eqnarray}
and the resulting order parameter corresponds to a magnetic octupole on a
triangular plaquette $\{ijk\}$.

As expected, both the real and the imaginary parts of the order
parameter transform with period $2\pi/3$
\begin{eqnarray}
    {\textstyle \mathcal{R}e \phantom. O^\dagger_{ijk}} &=& \phantom
    - {\textstyle {4 \over 3} ( O^{xxx}_{ijk} + R_{2\pi/3}^\sigma [O^{xxx}_{ijk}]
    + R_{4\pi/3}^\sigma [O^{xxx}_{ijk}] )},\nonumber\\
    {\textstyle \mathcal{I}m \phantom. O^\dagger_{ijk}}
    &=& - {\textstyle \frac{4}{3} ( O^{yyy}_{ijk}
    + R_{2\pi/3}^\sigma [O^{yyy}_{ijk}]
    + R_{4\pi/3}^\sigma [O^{yyy}_{ijk}] )},\nonumber
\end{eqnarray}
where $R_\theta^\sigma$ denotes a rotation in spin--space through
angle $\theta$ about $z$ axis.  The three--magnon condensate breaks
the remaining $U(1)$ symmetry down to the symmetries of a triangle
in the plane perpendicular to the applied field (Fig.~\ref{fig1}).
We therefore dub the resulting octupolar order ``triatic''.  We note
that triatic order of this type does {\it not} require any breaking
of translational symmetry --- the condensed three-magnon bound
states can move, as long as they remain phase coherent. Triatic
order of various types has previously been discussed in the context
of the Heisenberg AF on a kagom\'e lattice~\cite{kagome}.

Period--3 structure in the energy spectrum persists down to low spin
in all of the clusters studied. However, new features appear in the
structure of the spectrum for $S \lesssim 6$ (see
Fig.~\ref{fig:spectrum}). These features are sensitive to cluster
geometry, which suggests that different types of order compete with
octupolar order in the absence of magnetic field. While this makes
interpretation of the spectra more difficult, it is nonetheless
possible to place strong constraints on possible ground states.

First, it is possible to conclude that a new ground state exists
between the FM phase and the short-ranged RVB spin liquid phase with
finite spin-gap $\Delta \sim |J|$, which is known to occur for
$K/|J| \approx 0.5$~\cite{misguich98}. For $0.235 \lesssim K/|J|
\lesssim 0.28$ the ground state and low--lying excitations have
different symmetries from the RVB phase~\footnote{We also note that
the spectra showed no signature of the half--magnetization plateau
known to occur in the RVB phase~\cite{momoi97}.}. Furthermore, in
this parameter range, the rapid decay of excitation energies with
system size suggests a gapless ground state for $h=0$, and thus a
breaking of $SU(2)$ symmetry.

Dipolar (N\'eel), quadrupolar (nematic) and octupolar (e.g. triatic)
magnetic order all break $SU(2)$ symmetry, and each of these forms
of order is associated with a specific ``Anderson tower'' of
low--lying states~\cite{shannon06}.  The Anderson tower contains
states in all spin sectors $S$ for N\'eel order; only states with
even $S$ (period--2) for quadrupolar order, and only states in
sectors with $S$ a multiple of three for {\it pure} octupolar order
(period--3).

In view of the spectra, N\'eel order seems improbable. The number of
low--lying states below the continuum in each spin sector $S$ is not
compatible with what would be expected for dipolar order.
Furthermore, a period--2 structure where the lowest lying states
with odd $S$ are raised in energy relative to states with even $S$,
is seen for $S \lesssim 6$ in those $N=12$ and $N=36$ spin clusters
which possess the full symmetry of the lattice, and in the $N=24$
cluster with the lowest ground-state energy. This period--2
structure suggests quadrupolar order is present in the ground state.
The period--3 structure, present at larger $S$, is still clearly
visible for small $S$ in some $N=24$ clusters, but is hard to
distinguish in the lowest--energy clusters. Moreover, the ground
states for the $N=12$, $36$ clusters are not in the trivial irrep,
which indicate that the symmetry of the ground state differs from
the pure octupolar order observed in magnetic field.

\begin{figure}[tb]
\begin{center}
    \includegraphics[width=81mm]{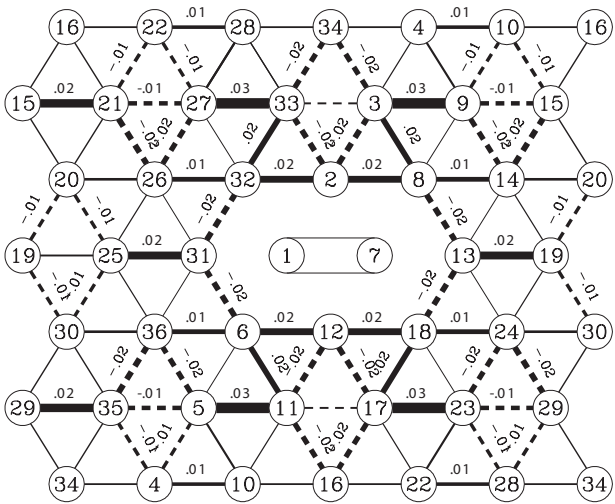}
\caption{Bond nematic correlation in the ground state of the $J$-$K$
model for $N=36$ spin cluster with $J = -4$, $K=1$. (See Ref.\
\cite{shannon06} for the definition.) Correlations are measured
relative to the reference bond \mbox{$1$--$7$}. Positive (negative)
correlations are drawn as full (dashed) lines. The thickness of the
lines is a measure of the strength of the correlation. }
\label{fig:correlation}
\end{center}
\end{figure}

In fact, the symmetry and momenta of the low lying even--$S$ states,
and the corresponding quadrupolar correlation function (shown in
Fig.~\ref{fig:correlation}), suggest the superposition of two
distinct types of quadrupolar order. One corresponds to the
condensation of s--wave magnon pairs with momenta at the $K$ points
(e.g. ${\bf q} = (0,2\pi/\sqrt{3})$) of the Brillouin zone
--- precisely the two--magnon instability of the saturated state predicted
to occur at $h_{c2}$. The other corresponds to
$d_{x^2-y^2}$+i$d_{xy}$--wave magnon pairs with momentum ${\bf q} =
(0,0)$.

We therefore conclude that the model has weak quadrupolar order in
its ground state which extends from $K/|J| \approx 0.23$ to $K/|J|
\approx 0.28$, where it undergoes a phase transition to a gapped RVB
phase~\footnote{It is also interesting to note that the gapful RVB
wave function contains states with crossed singlet/triplet structure
which is also seen in two--spin bound states near to
saturation~\cite{momoi05}.}. However in this parameter range, this
gapless nematic phase lies close to the boundary with other
competing phases.

We note that octupolar and quadrupolar orders are not exclusive:
octupolar order is known to be accompanied by quadrupolar order in
the ground state of spin $S=3/2$ models with polynomial
interactions~\cite{ChubukovB}. This raises the question of the
possibility of some similar coexistence in the present $S=1/2$
model. Short-range FM correlations on individual triangular
plaquettes (\mbox{$\langle ({\bf S}_1+{\bf S}_2+{\bf
S}_3)^2\rangle=2.42$} for $N=36$, $K/|J|=0.25$) persist within the
ground state, reflecting the tendency to form three--magnon bound
states. However, the present cluster sizes are too small to answer
this question.

The {\it gapless} spin liquid state observed in $^3$He films occurs
in the vicinity of the FM phase.
It seems reasonable that a nematic phase (or octupolar phase) should
interpolate between FM and RVB phases in the multiple spin exchange
model on the triangular lattice, and that this state is therefore a
strong contender to explain the spin--liquid seen in experiment.

It is our pleasure to acknowledge stimulating discussions with
P.~Fazekas, T.~Koretsune, K.~Kubo, K.~Penc and H.~Tsunetsugu. This
work was supported by MEXT Grants-in-Aid for Scientific Research
(No.\ 17071011 and No.\ 16GS0219), the Next Generation Super
Computing Project, Nanoscience Program, MEXT, Japan, and EPSRC grant
(EP/C539974/1). Large scale computations were performed at IDRIS
(Orsay).

\end{document}